\documentclass[12pt]{iopart}
\usepackage{graphicx}
\usepackage{tabularx}

\usepackage{amssymb}
\usepackage{color}
\usepackage{hyperref}
\usepackage{setstack}
\usepackage{bm}

\renewcommand{\vec}[1]{\bm{#1}}
\newcommand{\ket}[1]{|{#1} \rangle}
\newcommand{\bra}[1]{\langle {#1}|} 
\newcommand{\ip}[2]{\langle {#1} | {#2} \rangle}

\begin{document}

\title[Zone unfolding of Complex Bands]
{Brillouin zone unfolding of Complex Bands in a nearest neighbour Tight
Binding scheme}

\author{Arvind Ajoy$^1$, Kota V R M Murali$^2$,  and Shreepad
  Karmalkar$^1$}
\address{$^1$ Department of Electrical Engineering, Indian Institute of
Technology Madras, Chennai 600036, India}
\address{$^2$ IBM Semiconductor Research and Development Center,
Bangalore, India} 
\eads{\mailto{arvindajoy@iitm.ac.in},
      \mailto{kotamurali@in.ibm.com},
      \mailto{karmal@ee.iitm.ac.in}} 

\begin{abstract}
Complex bands $\vec{k}^{\perp}(E)$ in a semiconductor crystal, along a
general   direction   $\vec{n}$,    can   be   computed   by   casting
Schr\"odinger's  equation  as   a  generalized  polynomial  eigenvalue
problem.  When working with  primitive lattice  vectors, the  order of
this eigenvalue problem can grow large for arbitrary $\vec{n}$.  It is
however  possible to  always  choose a  set  of non-primitive  lattice
vectors  such  that  the   eigenvalue  problem  is  restricted  to  be
quadratic. The complex bands so  obtained need to be unfolded onto the
primitive Brillouin zone.  In this  paper, we present a unified method
to unfold real and complex bands.  Our method ensures that the measure
associated with  the projections of the  non-primary wavefunction onto
all candidate  primary wavefunctions is invariant with  respect to the
energy $E$.
\end{abstract}


\vspace{2pc}

\submitto{\JPCM}
\maketitle

\section{Introduction}
Complex  bandstructure $\vec{k}(E)$ describes  the properties  of both
propagating   and  evanescent   electronic  states   in  semiconductor
crystals.  Evanescent states have imaginary or complex wavevectors and
govern  tunneling   phenomena  \cite{Kane_JAP_1961}  in  semiconductor
devices.   The relative  importance of  these phenomena  has increased
with  every reduction  in the  dimensions of  these  devices.  Complex
bandstructure   is   also  used   to   predict   barrier  heights   of
metal-semiconductor   interfaces   \cite{Tersoff_PRL_1984}  and   band
lineups at semiconductor heterointerfaces \cite{Tersoff_PRB_1984}, via
the  theory  of  Virtual  Induced  Gap  States  (ViGS).   An  accurate
computation  of  complex  bandstructure  is hence  essential  for  the
continued  scaling and  materials engineering  of  electronic devices,
with an aim of improving performance.

\begin{figure}[t]
 \centering	
 \includegraphics[scale=1]{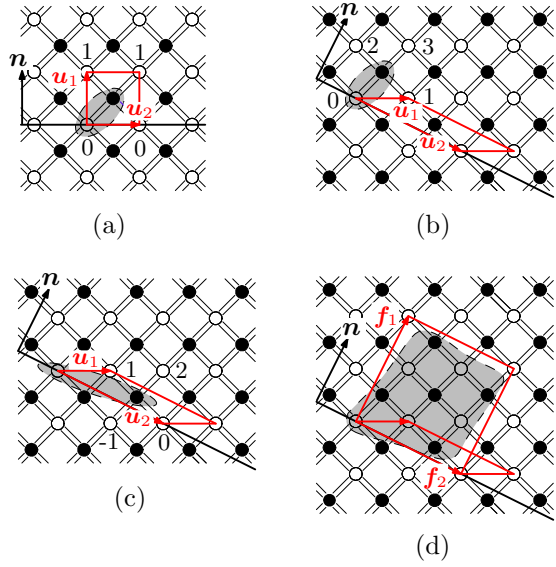}
 \caption{Two dimensional crystal  showing primitive and non-primitive
lattice  vectors  adapted  to  the  line  perpendicular  to  transport
direction $\vec{n}$.   The crystal  has a square  lattice and  a motif
consisting of one  $\circ$ (at $\vec{\nu} = 0$)  and one $\bullet$. In
each case, the motif is outlined by a dashed line and shaded gray. (a)
Primitive cell when $\vec{u}_1 \parallel \vec{n}$.  (b) Primitive cell
when $\vec{u}_1 \nparallel  \vec{n}$, using the same motif  as in (a).
(c) Primitive cell for the same  $\vec{n}$ as in (b), but with a motif
such  that  $\bullet$ is  within  the  cell.   (d) Non-primitive  cell
($\mathcal{N}_c = 5$) for the same  $\vec{n}$ as in (b). In cases (a),
(b), (c), the numbers indicate  the $s_1$ values (see (\ref{eq_rho}) )
of  the  nearest  neighbours  of  the  $\bullet$  of  the  motif.   The
corresponding  GPEP (\ref{eq_SchrodingerMatrixPrimitive}) is  of order
$\mathcal{O} = 2 \times \max(|s_1|)$. }
 \label{fig_TwoDimensionalCrystal}
\end{figure}

Of the many approaches  to bandstructure calculation, the $sp^3d^5s^*$
nearest neighbour empirical  tight binding method \cite{Jancu_PRB_1998,
Boykin_PRB_2004}  has proven  to  represent a  good trade-off  between
accuracy  and computational  efficiency. Complex  bands along  a given
transport direction $\vec{n}$ can be computed within this framework by
casting   Schr\"odinger's  equation   as   a  Generalized   Polynomial
Eigenvalue     Problem     (GPEP),     as    described     in    
\cite{Boykin_PRB_1996} for  the $[001]$ direction.   This method
can  be  extended  \cite{Ajoy_DRC_2011}  to a  general  $\vec{n}$,  by
working with a set of primitive lattice vectors $\vec{u}_1 , \vec{u}_2
, \vec{u}_3$ that are adapted to the plane perpendicular to $\vec{n}$,
i.e.  $\vec{u}_1 \cdot  \vec{n} > 0$ and $\vec{u}_2  , \vec{u}_3 \perp
\vec{n}$.   As  shown  in Figure  \ref{fig_TwoDimensionalCrystal}  and
described  in  Section  \ref{sec_Primitive},  the order  of  the  GPEP
depends on $\vec{n}$, since $\vec{u}_1$ is not necessarily parallel to
$\vec{n}$.

Hence, the  computation of complex bands along  an arbitrary $\vec{n}$
could  involve a GPEP  of large  order. Moreover,  arbitrary extrinsic
strain can  lead to  a GPEP  of large order  even for  transport along
simple directions  like $[111]$.  Robust  solution of a GPEP  of large
order  is  a  challenging \cite{Mackey_SIAM_2006}  problem,  sometimes
introducing large errors.  The order of  the GPEP can be limited to be
quadratic,   even  for   arbitrary  $\vec{n}$,   by  working   with  a
non-primitive set of  lattice vectors \cite{Laux_IWCE_2009} $\vec{f}_1
, \vec{f}_2  , \vec{f}_3$ such that $\vec{f}_1  \parallel \vec{n}$ and
$\vec{f}_2 ,  \vec{f}_3 \perp  \vec{n}$.  Energy bands  obtained using
this non-primitive cell correspond,  however, to primitive cell energy
bands that have been  folded onto the smaller, non-primitive Brillouin
zone.  These  bands have to  be unfolded onto the  primitive Brillouin
zone.

Zone folding and unfolding have  been studied extensively for the case
of    real    bands    \cite{Boykin_PRB_2005,    Boykin_EurJPhys_2006,
Boykin_JPhysCondMatter_2007,   Boykin_PRB_2007,   Boykin_Physica_2009,
Ku_PRB_2010}.  Computation  of real and complex bands  differ in their
choice   of  basis,   Bloch  sums   \cite{Slater_PhysRev_1954}  (which
represent the full periodicity of the lattice) for the former, whereas
Layer   Bloch  sums   \cite{Boykin_PRB_1996}  (which   only  represent
periodicity in directions perpendicular  to $\vec{n}$) for the latter.
Further,  unlike  wavefunctions  with  real  wavevectors,  those  with
complex wavevectors  need to  be normalized carefully.   The imaginary
part    of   the    wavevector   enters    into    the   normalization
constant. Ignoring this yields different  measures for the norm of the
wavefunction  for different $Im  (\vec{k})$. It  is hence  not obvious
whether the zone  unfolding method derived for real  bands can be used
to  unfold  complex  bands  along  a  general  $\vec{n}$.   Note  that
\cite{Laux_IWCE_2009} applies the scheme available for real bands to 
the case  of complex bands  without providing  any rigorous justification.
 
In this  paper, we  show rigorously that  the method of  unfolding can
indeed be used for complex  bands too, provided some modifications are
included. Our  modifications ensures that the  measure associated with
the  projections of  the non-primary  wavefunction onto  all candidate
primary wavefunctions is invariant with respect to the energy $E$, for
real and complex bands.   This invariance is especially important when
the supercell technique  \cite{Boykin_JPhysCondMatter_2007} is used to
compute the bandstructure of disordered materials.

This paper is structured  as follows.  In Section \ref{sec_Primitive},
we setup notation  and describe the method of  computing complex bands
along  a  general  $\vec{n}$  using plane  adapted  primitive  lattice
vectors.     Section   \ref{sec_NonPrimitive}    deals    with   using
non-primitive  vectors,  and  presents  the  modified  zone  unfolding
method.  Finally,  Section \ref{sec_Discussion} applies  our method by
to the  case of complex bands  along the $[110]$  direction in Silicon
and summarizes the paper.

\section{Complex bands using a primitive unit cell }
\label{sec_Primitive}

The  primitive  vectors   $\vec{u}_1$,  $\vec{u}_2$,  $\vec{u}_3$  are
constructed       using       the       method      described       in
\cite{Aravind_AmJPhys_2006}.  A point  in  the lattice  is
represented as
\begin{eqnarray}
\label{eq_rho}
\vec{\rho}(s_1,s_2,s_3) & = s_1\vec{u}_1 + 
\underbrace{s_2\vec{u}_2 + s_3\vec{u}_3}_
{\vec{\rho}^{\parallel}(s_2,s_3)} 
\end{eqnarray}
where  $s_1, s_2,  s_3$ are  integers.  Correspondingly,  a  vector in
reciprocal space is $\vec{k} = \vec{k}^{\parallel} + \vec{k}^{\perp}$,
such  that  $\vec{k}^{\perp}$  is  along $\vec{n}$.   The  crystal  is
constructed by associating  a motif of atoms with  each lattice point.
For crystals having a Zinc  Blende structure, the motif has two atoms.
Let $\vec{\nu}_{m}$, $m = 1,2$  represent the positions of these atoms
with respect  to the  lattice point.  We  set $\vec{\nu}_1  = \vec{0}$
without any loss of generality.

There are  $\mathcal{N}_{TB} =  20$ orthonormal orbitals  (10 L\"owdin
orbitals  \cite{Lowdin_JChemPhys_1950} of  each spin  type) associated
with each atomic site in  the $sp^3d^5s^*$ scheme.  An orbital of type
$\mu$, spin $\varsigma$ on an  atom $m$ located at site $\vec{\rho}_j$
is given by $\ip{\vec{r}}{\mu,\varsigma; \vec{\rho}_j + \vec{\nu}_m} =
\phi_{\mu\varsigma}\big(\vec{r} - (\vec{\rho}_j + \vec{\nu}_m)\big) $.
Complex   bands   are   obtained   by  expressing   the   wavefunction
$\psi(\vec{r},\vec{k})       =      \ip{\vec{r}}{\psi(\vec{k}^{\perp},
\vec{k}^{\parallel})}$  as a  linear combination  of layer  Bloch sums
\cite{Boykin_PRB_1996}. A layer Bloch sum is a linear superposition of
orbitals on all similar atoms  associated with a single lattice layer.
Denoting the layer Bloch sum  corresponding to orbital $\mu$ with spin
$\varsigma$  on atom $m$  in layer  $s_1 =  s$ as  $\xi_{\mu \varsigma
m}(\vec{r}; s, \vec{k}^{\parallel}) = \ip{\vec{r}}{\mu,\varsigma; m,s,
\vec{k}^{\parallel}}$, we have
\begin{eqnarray}
\label{eq_layerbloch_definition}
\xi_{\mu \varsigma m}(\vec{r};s, \vec{k}^{\parallel})  = 
\frac{1}{\sqrt{M_{\parallel}}} \sum_{j}^{(M_{\parallel})} 
e^{\iota \vec{k}^{\parallel} 
\cdot (\vec{\rho}^{\parallel}_j + s\vec{u}_1 + \vec{\nu}_m)} 
{\phi_{\mu\varsigma}(\vec{r}  - 
(\vec{\rho}^{\parallel}_j + s\vec{u}_1 + \vec{\nu}_m))}
\end{eqnarray}
where the  symbol $\sum_j^{(M_{\parallel})}$ denotes  a summation over
$M_{\parallel}$ lattice sites (indexed by $j$), within a parallelogram
with sides along  $\vec{u}_2, \vec{u}_3$. Periodic boundary conditions
are imposed  w.r.t this parallelogram. We thus write
\begin{eqnarray}
\label{eq_psi}
\ket{\psi(\vec{k}^{\perp}, \vec{k}^{\parallel})} =
\sum_{\mu \varsigma m} \sum_{s}^{(M_1)}
c^{\mu \varsigma m}_{s}(\vec{k}^{\perp}) 
\ket{\mu,\varsigma; m, s, \vec{k}^{\parallel}} 
\end{eqnarray}
as a summation over $M_1$ lattice layers. Both $M_{\parallel}$ and
$M_{1}$ are allowed to tend to infinity. 

The  periodicity of  the  lattice enforces  a  condition, 
\begin{eqnarray}
\label{eq_c}
c_{s}^{\mu \varsigma m}(\vec{k}^{\perp}) =
e^{\iota \vec{k}^{\perp} \cdot\vec{u}_1}
c_{s-1}^{\mu \varsigma  m}(\vec{k}^{\perp}). 
\end{eqnarray} 
Using  these layer  Bloch sums  as a  basis,  Schr\"odinger's equation
$\mathcal{H} \ket{\psi}  = E  \ket{\psi}$ can be  written as  a matrix
equation $\forall s$,
\begin{eqnarray}
\label{eq_SchrodingerMatrixPrimitive}
 \sum_{p}[{H}_{s,s-p}][{c}_{s-p}] + 
  \left([{H}_{s,s}] - [1]E \right)[{c}_{s}] 
 + \sum_{p}[{H}_{s,s+p}][{c}_{s+p}]  = 0
\end{eqnarray}
where $[c_s]$ is a column matrix of size $2 \mathcal{N}_{TB} \times 1$
such  that  $[c_s]_{\mu \varsigma  m}  =  c^{\mu  \varsigma m}_s$  and
$[H_{s',  s''}]$  is a  matrix  of  size  $2\mathcal{N}_{TB} \times  2
\mathcal{N}_{TB}$ such that
\begin{eqnarray}
\label{eq_HMatrixPrimitive}
 [H_{s',s''}]_{\mu'\varsigma' m', \mu''\varsigma'' m''} =   
\bra{\mu',\varsigma'; m', s', \vec{k}^{\parallel}}  
\mathcal{H}
\ket{\mu'',\varsigma''; m'', s'', \vec{k}^{\parallel}}.
\end{eqnarray}
The  summation in  (\ref{eq_SchrodingerMatrixPrimitive})  is over  all
unique $p \ne 0$ such that  the atom at $\vec{\rho}^{\parallel} + (s +
p) \vec{u}_1 +  \vec{\nu}_0$ is a nearest neighbour of  the atom at $s
\vec{u}_1 + \vec{\nu}_1$, for some $\vec{\rho}^{\parallel}$.  Equation
(\ref{eq_SchrodingerMatrixPrimitive})  is   a  generalized  polynomial
eigenvalue problem  (of order $\mathcal{O}  = 2\times \max  |p|$) with
eigenvalue $\lambda = e^{  \iota \vec{k}^{\perp} \cdot \vec{u}_1}$ and
eigenvector      $[c_s]$.       Following     \cite{Mackey_SIAM_2006},
(\ref{eq_SchrodingerMatrixPrimitive})  is said to  be $*$-palindromic,
since $[H_{s,s}]$  is Hermitian and  $[H_{s,s-p}]^{\dagger} = [H_{s-p,
s}] = [H_{s, s+p}]$; the  $\dagger$ refers to conjugate transpose. The
eigenvalues $\lambda$  thus occur in reciprocal  conjugate pairs, i.e.
if $\lambda$  is an eigenvalue, then $\frac{1}{\lambda^*}$  is also an
eigenvalue.   Hence, the component  of $\vec{k}^{\perp}(E)$  along the
transport  direction,  $k^{\perp}(E)$,   appears  in  conjugate  pairs
$(k^{\perp},             k^{\perp            *})$.            Equation
(\ref{eq_SchrodingerMatrixPrimitive}) can be solved for $k^{\perp}(E)$
by recasting  it as a generalized linear  eigenvalue problem involving
matrices of size $2  \cdot \mathcal{O} \cdot \mathcal{N}_{TB} \times 2
\cdot \mathcal{O} \cdot \mathcal{N}_{TB}$.

The    number   of    terms   and    the   order    $\mathcal{O}$   of
(\ref{eq_SchrodingerMatrixPrimitive})  depend  on  $\vec{n}$.  To  see
this,  consider the  toy two-dimensional  crystal as  shown  in Figure
\ref{fig_TwoDimensionalCrystal}.   This crystal  has a  square lattice
and a motif consisting of one $\circ$ and one $\bullet$.  Each $\circ$
is bonded  to four $\bullet$'s and  vice versa. The  numbers in Figure
\ref{fig_TwoDimensionalCrystal}(a),  (b), (c) give  the values  of $p$
required in (\ref{eq_SchrodingerMatrixPrimitive}) assuming $s = 0$.

\section{Complex bands using a Non-Primitive unit cell and a 
Modified Zone unfolding algorithm}
\label{sec_NonPrimitive}
Consider     the     non-primitive     unit     cell     in     Figure
\ref{fig_TwoDimensionalCrystal}(d).     Since   $\vec{f}_1   \parallel
\vec{n}$, atoms within  the motif bond to atoms  belonging only to the
same or neighbouring  lattice layers.  Thus, in the general  case, we can ensure
that the generalized polynomial eigenvalue problem is restricted to be
quadratic, by working  with non-primitive vectors $\vec{f}_1 \parallel
\vec{n},\vec{f}_2 = \vec{u}_2, \vec{f}_3  = \vec{u}_3$.  The volume of
the non-primitive unit cell is an integral multiple $\mathcal{N}_c$ of
that of  the primitive cell,  causing the the  non-primitive Brillouin
zone  to be $1/\mathcal{N}_c$  as large  as the  primitive one.  As an
example,        $\mathcal{N}_c       =       5$        in       Figure
\ref{fig_TwoDimensionalCrystal}(d).  We  choose the non-primitive unit
cell to have the same origin as the primitive cell.  We use upper case
Roman  and   Greek  letters  to  denote  quantities   related  to  the
non-primitive scheme.

A non-primitive  lattice point is given by
\begin{eqnarray}
\label{eq_R}
\vec{R}(t_1,t_2,t_3) & = t_1\vec{f}_1 + 
\underbrace{t_2\vec{f}_2 + t_3\vec{f}_3}_
{\vec{R}^{\parallel}(t_2,t_3)} 
\end{eqnarray}
where $t_1, t_2,  t_3$ are integers.  A vector  in reciprocal space is
now  $\vec{K}  = \vec{K}^{\parallel}  +  \vec{K}^{\perp}$.  The  motif
associated with  each lattice point will  have $2\mathcal{N}_c$ atoms,
positioned at  $\vec{\gamma}_{n}$, $n = 1,  2, \ldots, 2\mathcal{N}_c$
w.r.t the  lattice point. Since the primitive  and non-primitive cells
share  a  common  origin,  we  set  $\vec{\nu}_1  =  \vec{\gamma}_1  =
\vec{0}$.  We   denote  the  non-primitive  layer   Bloch  sum  (over
$N_{\parallel}$ lattice  sites) as $\Xi_{\mu  \varsigma n}(\vec{r}; t,
\vec{K}^{\parallel})      =      \ip{\vec{r}}{\mu,\varsigma;      n,t,
\vec{K}^{\parallel}}  $ and  wavefunction as  $\Psi(\vec{r},\vec{K}) =
\ip{\vec{r}}{\Psi(\vec{K}^{\perp}, \vec{K}^{\parallel})}$. Writing
\begin{eqnarray}
\label{eq_Psi}
\ket{\Psi(\vec{K}^{\perp}, \vec{K}^{\parallel})} =
\sum_{\mu \varsigma n} \sum_{t}^{(N_1)}
C^{\mu \varsigma n}_{t}(\vec{K}^{\perp}) 
\ket{\mu,\varsigma; n, t, \vec{K}^{\parallel}},
\end{eqnarray}
we  obtain   $K^{\perp}(E)$  by  solving   the  resulting  generalized
quadratic eigenvalue problem. Finally, $k^{\perp}(E)$ is computed from
$K^{\perp}(E)$ using  the modified zone  unfolding algorithm described
below. 

It  is important to  recognize that  working with  large non-primitive
cells  could present  numerical difficulties  in the  solution  of the
generalized quadratic eigenvalue problem. Poor quality eigenvalues and
eigenvectors  could render  the zone  unfolding method  useless.  This
problem is expected to be most severe for eigenvalues corresponding to
large $|Im(K^{\perp})|$, owing to the exponential nature of the factor
$\lambda =  e^{ \iota \vec{K}^{\perp} \cdot  \vec{f}_1}$. However, the
problem  is  mitigated  by  the  fact that  our  primary  application,
modelling  of  tunneling phenomena,  only  requires evanescent  states
having  the   smallest  $|Im(K^{\perp})|$.   Nevertheless,   the  most
important reason for erroneous eigenvalues and eigenvectors is the use
of      the      standard      companion     linearization      scheme
\cite{Mackey_SIAM_2006}, which  neglects the palindromic  structure of
the  GPEP (as shown,  for example,  in \cite{Ipsen_SIAM_2004}  for the
case  of vibration analysis  of fast  trains, involving  an eigenvalue
problem  with similar  symmetry). The  eigenvalues $\lambda$  hence no
longer appear  as $\lambda, \frac{1}{\lambda^*}$ pairs.  The  use of a
structure     preserving     linearization     \cite{Mackey_SIAM_2006,
Mackey_SIAM_2006_2,  Huang_NumerMath_2011} rectifies  this  issue, and
has been shown  to greatly improve the quality  of the eigenvalues and
eigenvectors. Thus, a careful  choice of linearization and eigensolver
is critical to  the scalability of the method  discussed in this paper
to large non-primitive cells.

The  essential idea  in zone  unfolding is  to express  a wavefunction
obtained  using  a  non-primitive  cell  as a  linear  combination  of
primitive cell wavefunctions. The process of unfolding then boils down
to  estimating  the contributions  of  each  of  these primitive  cell
wavefunctions  to the  non-primitive cell  wavefunction.  In  order to
achieve this,  both the non-primitive and  primitive wavefunctions are
written in terms of their  constituent atomic orbitals.

\subsection{Wavefunctions in terms of atomic orbitals}
\label{sec_WavefnIntermsofAtomicOrbitals}
To remain consistent with the  zone unfolding algorithm for real bands
available     in    \cite{Boykin_JPhysCondMatter_2007,
Boykin_PRB_2007},  we use  a slightly  modified version  of  the layer
Bloch sums to describe the zone unfolding procedure.  Working with the
non-primitive  cell, we  define a  primed layer  Bloch  sum $\Xi'_{\mu
\varsigma         n}(\vec{r};t,         \vec{K}^{\parallel})         =
\ip{\vec{r}}{\mu,\varsigma; n,t, \vec{K}^{\parallel}}'$,
\begin{eqnarray}
\label{eq_LayerBloch_definition_primed}
\Xi'_{\mu  \varsigma n}(\vec{r};t,  \vec{K}^{\parallel})  = 
\frac{1}{\sqrt{N_{\parallel}}} \sum_{j}^{(N_{\parallel})} 
e^{\iota \vec{K}^{\parallel} 
\cdot (\vec{R}^{\parallel}_j + t\vec{f}_1)} 
\ip{\vec{r}}{\mu,\varsigma; \vec{R}^{\parallel}_j + t\vec{f}_1 +
  \vec{\gamma}_n }
\end{eqnarray} 
Notice that  this differs  from the non-primary  version of  the layer
Bloch  sum  defined in  (\ref{eq_layerbloch_definition})  only in  the
absence    of   the    term   $e^{\iota    \vec{K}^{\parallel}   \cdot
\vec{\gamma}_n}$    preceding   the    atomic    orbital.    Following
(\ref{eq_Psi}), we write
\begin{eqnarray}
\label{eq_PsiPrimed}
\ket{\Psi(\vec{K}^{\perp}, \vec{K}^{\parallel})} =
\sum_{\mu \varsigma n} \sum_{t}^{(N_1)}
C'^{\; \mu \varsigma n}_{t}(\vec{K}^{\perp}) \ket{\mu,\varsigma; n, t, \vec{K}^{\parallel}}', 
\end{eqnarray}
where, we have similar to (\ref{eq_c}),
\begin{eqnarray}
\label{eq_CPrimed}
C_{t+1}'^{\; \mu \varsigma n}(\vec{K}^{\perp}) = 
e^{\iota \vec{K}^{\perp} \cdot\vec{f}_1}
C_t'^{\; \mu \varsigma  n}(\vec{K}^{\perp})
\end{eqnarray}
Comparing the two expansions for the wavefunction (\ref{eq_Psi}),
(\ref{eq_PsiPrimed}) we can relate the expansion coefficients in the
primed basis to those obtained in the unprimed basis as
\begin{eqnarray}
\label{eq_C_CPrimed}
C'^{\;\mu \varsigma n}_{t}(\vec{K}^{\perp})  =
e^{\iota  \vec{K}^{\parallel}  \cdot   \vec{\gamma}_n  }
C^{\mu \varsigma n}_{t}(\vec{K}^{\perp}). 
\end{eqnarray}

We now attempt to rewrite  the expansion (\ref{eq_PsiPrimed}) in a way
such that the condition  (\ref{eq_CPrimed}) is explicitly imposed. For
this, we introduce a quantity  $\tilde{C}'^{\; \mu \varsigma n}$ which is
independent of layer $t$, such that
\begin{eqnarray}
\label{eq_CPrimeTilde}
C_t'^{\; \mu \varsigma n}(\vec{K}^{\perp}) = 
\frac{e^{\iota  \vec{K}^{\perp} \cdot t\vec{f}_1 }}{\sqrt{S_{NP}(\vec{K}^{\perp})}} 
\tilde{C}'^{\mu \varsigma n} (\vec{K}^{\perp}),
\end{eqnarray}
where  $S_{NP}(\vec{K}^{\perp})$  is  a  normalization  constant  (the
subscript  $NP$   refers  to  non-primitive).  From
(\ref{eq_LayerBloch_definition_primed}), (\ref{eq_PsiPrimed}),
(\ref{eq_CPrimeTilde}), we have
\begin{eqnarray}
\label{eq_Psi:1}
\nonumber \fl \ket{\Psi(\vec{K}^{\perp}, \vec{K}^{\parallel})}  
= \frac{1}{\sqrt{N_{\parallel}S_{NP}(\vec{K}^{\perp})}} 
& \sum_{\mu \varsigma n}\sum_{j}^{(N_{\parallel})}\sum_{t = 0}^{N_1 - 1} 
  \tilde{C}'^{\; \mu \varsigma n}(\vec{K}^{\perp})  \times \\ 
& e^{\iota  \vec{K}^{\perp} \cdot t\vec{f}_1 }
e^{\iota \vec{K}^{\parallel} 
\cdot (\vec{R}^{\parallel}_j + t\vec{f}_1)}
\ket{\mu,\varsigma; \vec{R}^{\parallel}_j + t\vec{f}_1 +  \vec{\gamma}_n} 
\end{eqnarray}
Note that we  have explicitly chosen the limits $t =  0, \ldots, N_1 -1$
for the sum $\sum_{t}^{(N_1)}$. The  reason for this will become clear
in  Section   \ref{sec_RelationshipBetPrimitiveNonPrimitive}  when  we
consider   the  relationship   between  non-primitive   and  primitive
reciprocal  vectors.  In  short,  we  wish to  ensure  that the  atoms
considered  when working  with  non-primitive or  primitive cells  are
identical.

One can use the fact that $\vec{K}^{\perp} \cdot \vec{R}^{\parallel}_j
= 0$ and simplify the exponent in  (\ref{eq_Psi:1}) as 
\begin{eqnarray}
\label{eq_Kparallel.Kperp:K}
\vec{K}^{\perp} \cdot t\vec{f}_1  + 
\vec{K}^{\parallel} \cdot (\vec{R}^{\parallel}_j + t\vec{f}_1) = 
(\vec{K}^{\perp} + \vec{K}^{\parallel}) \cdot (\vec{R}^{\parallel}_j + t\vec{f}_1) 
= \vec{K} \cdot \vec{R}_{j'},
\end{eqnarray}
where $\vec{R}_{j'} = (\vec{R}^{\parallel}_j + t\vec{f}_1)$.
Hence,  using (\ref{eq_Kparallel.Kperp:K})  to  rewrite the  double
summation  in (\ref{eq_Psi:1}), $\sum_{j}^{(N_{\parallel})}  \sum_{t =
0}^{N_1    -   1}   \equiv    \sum_{j'}^{(\mathcal{N}_{NP})}$   (where
$\mathcal{N}_{NP}  =  N_{\parallel}N_{1}$  refers  to  the  number  of
non-primitive lattice points) and dropping the $'$ on $j'$, we get
\begin{eqnarray}
\label{eq_Psi:2}
\nonumber \fl \ket{\Psi(\vec{K}^{\perp}, \vec{K}^{\parallel}}  =
\sqrt{\frac{N_1}{S_{NP}(\vec{K}^{\perp})}}\frac{1}{\sqrt{\mathcal{N}_{NP}}} 
\sum_{\mu \varsigma n}\sum_{j}^{(\mathcal{N}_{NP})}
\tilde{C}'^{\; \mu \varsigma n}(\vec{K}^{\perp}) \times 
e^{\iota  \vec{K} \cdot \vec{R}_j}
\ket{\mu,\varsigma; \vec{R}_j + \vec{\gamma}_n}
\end{eqnarray}
We  have   thus  been  able   to  rewrite  $\ket{\Psi(\vec{K}^{\perp},
\vec{K}^{\parallel}}$ in terms of  the full $\vec{K} = \vec{K}^{\perp}
+  \vec{K}^{\parallel}$.   Provided we  have  $\sum_{\mu \varsigma  n}
|\tilde{C}'^{\;  \mu   \varsigma  n}(\vec{K}^{\perp})|^2  =   1$,  the
expression  (\ref{eq_Psi:2}) is very  similar to  the one  employed in
 \cite{Boykin_JPhysCondMatter_2007} for  the  case of  real
bands,         except         for         the        factor         of
$\sqrt{\frac{N_1}{S_{NP}(\vec{K}^{\perp})}}$.   Indeed,  this  is  the
reason that the zone unfolding  procedure developed for real bands can
be  applied to  the  case of  complex  bands, albeit  with some  minor
modifications.

We  can now  write out  an expression  for the  normalization constant
$S_{NP}(\vec{K}^{\perp})$  so that  wavefunction  is normalized,  i.e.
$\ip{\Psi(\vec{K}^{\perp},                        \vec{K}^{\parallel})}
{\Psi(\vec{K}^{\perp},   \vec{K}^{\parallel})}  =1$,   and  $\sum_{\mu
\varsigma  n} |\tilde{C}'^{\;  \mu \varsigma  n}(\vec{K}^{\perp})|^2 =
1$.     Note    that    $\vec{K}^{\parallel}$   is    real;    however
$\vec{K}^{\perp}$ can be complex  in general.  Using the orthogonality
of the L\"owdin orbitals, we get
\begin{eqnarray}
\label{eq_S_NP}
\nonumber S_{NP}(\vec{K}^{\perp}) &= \sum_{t = 0}^{N_1 - 1}
e^{-t \alpha }, 
\mathrm{ where } \; \alpha = 2 Im(\vec{K}^{\perp} \cdot \vec{f}_1) \\
& =  \cases{
  N_1,  
  & if $\alpha = 0$, \\
  \frac{1 - e^{-\alpha N_1}}{1 - e^{-\alpha}}, 
  & if $\alpha  \ne 0$.
  }
\end{eqnarray}
Note that $S_{NP}(\vec{K}^{\perp}) = N_1$ irrespective of the value of
$\vec{K}^{\perp}$ when the energy of the wavefunction corresponds to
a real band (i.e $\alpha = 0$). However, $S_{NP}(\vec{K}^{\perp})$
depends on $Im (\vec{K}^{\perp})$ in general.

As described in the appendix, the   primitive  wavefunction is recast
similarly as
\begin{eqnarray}
\label{eq_psi:2}
\ket{\psi(\vec{k}^{\perp}, \vec{k}^{\parallel})} = 
\sqrt{\frac{M_1}{S_{P}(\vec{k}^{\perp})}}\frac{1}{\sqrt{\mathcal{N}_{P}}} 
\sum_{\mu \varsigma m}\sum_{j}^{(\mathcal{N}_{P})}
\tilde{c}'^{\; \mu \varsigma m}(\vec{k}^{\perp})
e^{\iota  \vec{k} \cdot \vec{\rho}_j}
\ket{\mu,\varsigma; \vec{\rho}_j + \vec{\nu}_m}
\end{eqnarray}
with   $S_{P}(\vec{k}^{\perp})$   being   a  normalization   constant.
$\mathcal{N}_{P}  =  M_{\parallel}M_{1}$   refers  to  the  number  of
primitive lattice points. $\tilde{c}'^{\; \mu \varsigma m}$ is related
to  the expansion  coefficients $c^{\mu  \varsigma m}_s$  by equations
similar  to  (\ref{eq_C_CPrimed}),  (\ref{eq_CPrimeTilde}).

\subsection{Relationship between primitive and
    non-primitive reciprocal vectors}
\label{sec_RelationshipBetPrimitiveNonPrimitive}
By construction, $\vec{u}_1$  and $\vec{f}_1$ lie to the  same side of
the plane perpendicular to $\vec{n}$.  Since the lattice points in the
non-primitive lattice are a subset  of those in the primitive lattice,
the ratio $\frac{\vec{n} \cdot \vec{f}_1 }{\vec{n} \cdot \vec{u}_1 } =
L_1$  is  an integer.  Again,  as  an example,  $L_1  =  5$ in  Figure
\ref{fig_TwoDimensionalCrystal}(d).    Physically,    there   are   $L_1$
primitive lattice layers within  a single non-primitive lattice layer.
Thus the  non-primitive and primitive  surface adapted unit  cells are
commensurate   \cite{Boykin_EurJPhys_2006}  with   each   other  along
$\vec{n}$.  On  the other hand, the non-primitive  and primitive cells
are  not necessarily  commensurate within  the plane  perpendicular to
$\vec{n}$.  Since  the non-primitive cell is  $\mathcal{N}_c$ times as
large  as  the primitive  cell,  $\mathcal{N}_c$ primitive  reciprocal
vectors $\vec{k}_{\theta},  \theta =  1, 2, \ldots  \mathcal{N}_c$ map
onto  the  same non-primitive  reciprocal  vector  $\vec{K} $.   Using
results  available in  \cite{Boykin_EurJPhys_2006},  we can
write
\begin{eqnarray}
\label{eq_kKq}
\vec{k}_{\theta} = \vec{K} + \vec{q}_{\theta}, \quad
\theta =  1, 2, \ldots  \mathcal{N}_c,
\end{eqnarray}
where $\vec{q}_{\theta}$ is a  vector in the first primitive Brillouin
zone  (and  hence purely  real)  that  is  commensurate with  periodic
boundary conditions on the non-primitive cell, i.e.
\begin{eqnarray}
\label{eq_q_definition}
\vec{f}_i \cdot \vec{q}_{\theta} = 2 \pi \times \mathtt{integer},
\quad i = 1,2,3.
\end{eqnarray}
Note that  $\forall \theta', \theta''  = 1, 2,  \ldots \mathcal{N}_c$,
$\theta' \ne  \theta''$, no  $\vec{q}_{\theta'}$ should be  related to
any  other  $\vec{q}_{\theta''}$  by  a primitive  reciprocal  lattice
vector.

We can now justify the choice of summation limits used for $t$, $s$ in
(\ref{eq_Psi:1}), (\ref{eq_psi:1})  respectively. First, we  point out
that  the layer Bloch  sums remain  invariant upon  a shift  in atomic
position, by a lattice vector  in the plane perpendicular to $\vec{n}$
(i.e.   $\forall$ integers  $\alpha_2,  \alpha_3$,  a  shift  $
\alpha_2\vec{u}_2 +  \alpha_3\vec{u}_3$ in  the case of  primitive and
$\alpha_2\vec{f}_2 +  \alpha_3\vec{f}_3$ in the  case of non-primitive
layer Bloch sums). Such a shift  merely refers to an identical atom in
the motif  at a different  lattice site.  This invariance  arises from
the  summation over all  lattice sites,  given that  periodic boundary
conditions  are  implied  on   the  boundaries  of  the  parallelogram
enclosing the lattice sites.   However, no periodic boundary condition
can  be applied  along $\vec{u}_1,  \vec{f}_1$ when  the perpendicular
component of the reciprocal vector is complex. Now, let us choose $N_1
=  L_1 M_1$.   Then, $t  = 0,  1, \ldots,  (N_1 -1)$  and $s  =  0, 1,
\ldots,(M_1    -1)$   ensures    that   sums    in   (\ref{eq_Psi:1}),
(\ref{eq_psi:1})  run over the  same physical  space in  the $\vec{n}$
direction; in fact, one could in  general choose $t = t', \ldots, (N_1
-1 +  t')$ and $s =  L_1t', \ldots,(M_1 -1 +  L_1t')$. Further, assume
that the  primitive motif is such  that its atoms are  within the unit
cell         (for          example,         as         in      Figure
\ref{fig_TwoDimensionalCrystal}(c)).   Then,   the   sets   of   atoms
considered when  working with non-primitive or  primitive cells differ
in  position  only by  some  $\alpha_2\vec{f}_2 +  \alpha_3\vec{f}_3$,
which, in the light of the above discussion implies that the atoms are
identical.   This  is  important  when we  express  the  non-primitive
wavefunction as a linear combination of primitive wavefunctions.

\subsection{Non-primitive wavefunction in terms of
  primitive wavefunctions}
\begin{figure}[t]
 \centering	
 \includegraphics[scale=1]{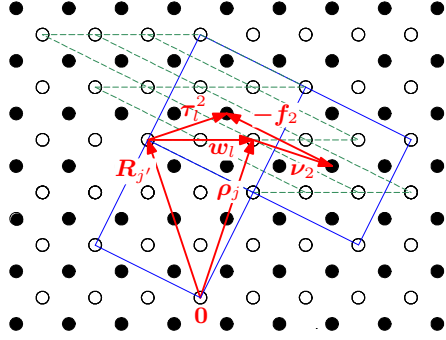}
 \caption{Remapping   of   atomic  positions   in   the  primitive   and
non-primitive  descriptions,  shown   for  one  particular  $\bullet$
$(m=2)$     of    the    two     dimensional    crystal     of    Fig.
\ref{fig_TwoDimensionalCrystal}.    Primary   motif   corresponds   to
Fig. \ref{fig_TwoDimensionalCrystal}(c). Thin dashed and solid lines
show a few primitive and non-primitive cells respectively. Bonds
between $\circ$ and $\bullet$ not shown, for clarity.}
 \label{fig_AtomRemap}
\end{figure}

Consider  a   non-primitive  wavefunction  $\ket{\Psi(\vec{K})}$  with
energy      $E$.      Using     (\ref{eq_kKq}),      and     following
\cite{Boykin_PRB_2005}, we express $\ket{\Psi(\vec{K})}$ in
terms  of primitive wavefunctions  $\ket{\psi(\vec{k}_{\theta})}$ that
have the same energy E. Thus,
\begin{eqnarray}
\label{eq_Psi_psi}
\ket{\Psi(\vec{K})} =
\sum_{\theta = 1}^{\mathcal{N}_c}a_{\theta}
\ket{\psi(\vec{k}_{\theta})} 
= \sum_{\theta = 1}^{\mathcal{N}_c}a_{\theta}
\ket{\psi(\vec{K} + \vec{q}_{\theta})}.
\end{eqnarray}
The  motif   associated  with   a  non-primitive  lattice   point  has
$2\mathcal{N}_c$  atoms.   The  primitive  motif has  $2$  atoms.   We
introduce  $\vec{\tau}_l^m  =  \vec{\gamma}_n$,  with $l=  1,  \ldots,
\mathcal{N}_c$ and  $m = 1,2$ to  denote the position  of the $l^{th}$
atom of type $m$ (w.r.t  the primitive motif) within the non-primitive
motif. Correspondingly, $\tilde{C}'^{\mu  \varsigma n}$ can be designated
as  $\tilde{C}'^{\;  \mu  \varsigma  m  l}$.  Hence  (\ref{eq_Psi:2})  is
modified as
\begin{eqnarray}
\label{eq_Psi:final}
\nonumber \fl \ket{\Psi(\vec{K})}  =
\sqrt{\frac{N_1}{S_{NP}(\vec{K}^{\perp})}}\frac{1}{\sqrt{\mathcal{N}_{NP}}} 
\sum_{\mu \varsigma m}
\sum_{j}^{(\mathcal{N}_{NP})} 
&\sum_{l = 1}^{\mathcal{N}_c} 
\tilde{C}'^{\; \mu \varsigma m l}(\vec{K}^{\perp}) \times \\
&e^{\iota  \vec{K} \cdot \vec{R}_j}
\ket{\mu,\varsigma; \vec{R}_j + \vec{\tau}_l^m}.
\end{eqnarray}
Further,  each non-primitive  unit cell  will  enclose $\mathcal{N}_c$
primitive    lattice    points.     Let    $\vec{w}_l,    \;    l    =
1,\ldots,\mathcal{N}_c$  denote  the   positions  of  these  primitive
lattice points within a non-primitive cell, with respect to the common
origin of the primitive  and non-primitive cells. In (\ref{eq_psi:2}),
one  can map  the  atomic positions  $\vec{\rho}_j  + \vec{\nu}_m$  to
equivalent  atomic  positions  $\vec{R}_{j'} +  \vec{\tau}_l^m$  where
$\vec{\rho}_{j}  =  \vec{R}_{j'} +  \vec{w}_l$  and $\vec{\tau_l^m}  =
\vec{w}_l   +  \vec{\nu}_m   +  s_2\vec{f}_2   +   s_3\vec{f}_3$  (the
equivalence as  explained previously is established  for some integers
$s_2, s_3$ ).  Figure  \ref{fig_AtomRemap} represents  the above mapping
pictorially      using     the     two-dimensional      crystal     of
Figure         \ref{fig_TwoDimensionalCrystal}.          The         sum
$\sum_{j}^{\mathcal{N}_{P}}$  can then  be  replaced by  a double  sum
$\sum_{j'}^{\mathcal{N}_{NP}}  \sum_{l= 1}^{\mathcal{N}_c}$.  Dropping
the $'$ on $j'$, and including (\ref{eq_kKq}), we thus get from (\ref{eq_psi:2}),
\begin{eqnarray}
\label{eq_psi_Kq}
\nonumber \fl \ket{\psi(\vec{K} + \vec{q}_{\theta})}  = 
\sqrt{\frac{M_1}{S_{P}((\vec{K} + \vec{q}_{\theta})^{\perp})}}
&\frac{1}{\sqrt{\mathcal{N}_{NP}\mathcal{N}_c}} 
\sum_{\mu \varsigma  m}
\sum_{j}^{(\mathcal{N}_{NP})}
\sum_{l=1}^{\mathcal{N}_c} \\
&\tilde{c}'^{\; \mu \varsigma m}((\vec{K} + \vec{q}_{\theta})^{\perp})
e^{\iota  (\vec{K} + \vec{q}_{\theta}) \cdot (\vec{R}_j + \vec{w}_l)}
\ket{\mu,\varsigma; \vec{R}_j + \vec{\tau}_l^m}.
\end{eqnarray}
We   now  substitute   (\ref{eq_Psi:final})  and   (\ref{eq_psi_Kq})  in
(\ref{eq_Psi_psi}).          Note         that         $S_{P}((\vec{K}
+\vec{q}_{\theta})^{\perp})    =   S_{P}(\vec{K}^{\perp})$    and   is
independent   of   $\theta$,   since  $\vec{q}_{\theta}$   is   purely
real. Additionally,  $\vec{q}_{\theta} \cdot \vec{R}_j =  2 \pi \times
\mathtt{integer}$ from (\ref{eq_R}), (\ref{eq_q_definition}).  We then
compare   the  coefficients   of   $\ket{\mu,\varsigma;  \vec{R}_j   +
\vec{\tau}_l^m}$ on both sides of (\ref{eq_Psi_psi}).  Rearranging the
terms,  we  obtain a  system  of  $\mathcal{N}_c$  equations for  each
combination $\mu \varsigma m$,
\begin{eqnarray}
\label{eq_semifinal_1}
e^{-\iota \vec{K} \cdot \vec{w}_l} 
\tilde{C}^{' \; \mu \varsigma  m l}(\vec{K}^{\perp}& )  = 
\frac{\Lambda}{\sqrt{\mathcal{N}_c}}
\sum_{\theta = 1}^{\mathcal{N}_c}
e^{\iota \vec{w}_l \cdot \vec{q}_{\theta}} \times  
a_{\theta}
\tilde{c}'^{\; \mu \varsigma m}((\vec{K} +\vec{q}_{\theta})^{\perp}) 
\end{eqnarray}
where $l  =  1, \dots, \mathcal{N}_c$ and 
\begin{eqnarray}
\label{eq_semifinal_2}
\Lambda  =   \sqrt {\frac{M_1}{N_1}
\frac{S_{NP}(\vec{K}^{\perp})}{S_{P}(\vec{K}^{\perp})} }.
\end{eqnarray}

In  order to  simplify  $\Lambda$, we  point  out that  (\ref{eq_kKq})
implies  $Im(\vec{k_{\theta}}^{\perp})  = Im(\vec{K}^{\perp})$.  Hence
$\alpha   =  2Im(\vec{K}^{\perp}   \cdot  \vec{f}_1)   =   L_1  \times
2Im(\vec{k_{\theta}}^{\perp} \cdot \vec{u}_1) = L_1 \beta$. Also note
that $M_1 = L_1 N_1$. Thus, from (\ref{eq_S_NP}), (\ref{eq_S_P}) we have
\begin{eqnarray}
\label{eq_Lambda}
\Lambda = 
\cases{ 1, & if  $\alpha = \beta = 0$, \\
        \sqrt{L_1  \frac{1 - e^{-\beta}}{1 - e^{-\alpha}}},  
           & if  $ \alpha, \beta \ne 0$ .
      }     
\end{eqnarray}
It is important to appreciate that  our choices of $M_1 = L_1 N_1$ and
summation  limits  for  $t,s$  in  (\ref{eq_Psi:1}),  (\ref{eq_psi:1})
ensure that though $S_{NP}, S_{P} \to \infty$ as $N_1, M_1 \to \infty$
and $\alpha,  \beta < 0$, the ratio  $\frac{S_{NP}}{S_{P}}$ is always
well behaved.

We can transform (\ref{eq_semifinal_1}) into a matrix equation,
\begin{eqnarray}
\label{eq_final}
[B_{\mu \varsigma m}] &=& 
\Lambda \; [U] \cdot [A_{\mu \varsigma  m}]
\end{eqnarray}
where 
\begin{eqnarray*}
\label{B}
{[B_{\mu \varsigma m}]} = &
  \left[  
  \begin{array}{c}
    e^{- \iota \vec{K} \cdot \vec{w}_1} 
    \tilde{C}'^{\;\mu \varsigma m 1}(\vec{K}^{\perp})\\
    \vdots \\
    e^{- \iota \vec{K} \cdot \vec{w}_{\mathcal{N}_c}} 
    \tilde{C}'^{\;\mu \varsigma m \mathcal{N}_c}(\vec{K}^{\perp})
  \end{array}
  \right], \\
\label{eq_U}
{[U]} = 
  \frac{1}{\sqrt{\mathcal{N}_c}}&
  \left[
  \begin{array}{ccc}
    e^{\iota \vec{w}_1 \cdot \vec{q}_1} &    
    \cdots &
    e^{\iota \vec{w}_1 \cdot \vec{q}_{\mathcal{N}_c}} \\
    \vdots & \ddots  & \vdots \\
    e^{\iota \vec{w}_{\mathcal{N}_c} \cdot \vec{q}_1} &    
    \cdots &
    e^{\iota \vec{w}_{\mathcal{N}_c} \cdot \vec{q}_{\mathcal{N}_c}} \\
  \end{array}
  \right], \\
\label{A}
{[A_{\mu \varsigma  m}}] = &
 \left[
 \begin{array}{c}
    a_1 \tilde{c}'^{\; \mu \varsigma m}((\vec{K} + \vec{q}_1)^{\perp}) \\
    \vdots \\
    a_{\mathcal{N}_c} 
    \tilde{c}'^{\; \mu \varsigma m}((\vec{K} + \vec{q}_{\mathcal{N}_c})^{\perp})
  \end{array}
 \right] .  
\end{eqnarray*}
We  remark  that (\ref{eq_final})  is  very  similar  to the  equation
derived  in  \cite{Boykin_PRB_2005}  for the  case of  real
bands,   except  for  the   additional  factor   $\Lambda$.  Following
 \cite{Boykin_PRB_2005}, we solve (\ref{eq_final}) to obtain
$[A_{\mu  \varsigma m}]$ for  all combinations  of $\mu  \varsigma m$,
using  the property  that $[U]$  is  unitary.  Since  we have  ensured
$\sum_{\mu \varsigma  m}|\tilde{c}'^{\; \mu \varsigma m}|^2 =  1 $, we
obtain
\begin{eqnarray}
  a_{\theta} = \sqrt{\sum_{\mu \varsigma m} {\Big|[A_{\mu \varsigma m}]_{\theta}\Big|}^2}
\end{eqnarray}
Further,  since both  $\ket{\Psi(\vec{K})}$  and $\ket{\psi(\vec{k})}$
have  been normalized,  and wavefunctions  corresponding  to different
wavevectors   are  orthogonal,   the  measure   associated   with  the
projections  of the  non-primitive  wavefunction $\ket{\Psi(\vec{K})}$
onto   candidate    primitive   wavefunctions   $\ket{\psi(\vec{K}   +
\vec{q}_{\theta})}$,
\begin{eqnarray}
\label{a_measure}
 \mathcal{M} = \sum_{\theta = 1}^{\mathcal{N}_c}|a_{\theta}|^2   =  1,
\end{eqnarray}
independent of  the energy $E$. It  is reasonable to  expect that most
$a_{\theta}$  will  be  zero.  The primitive  wavevectors  $\vec{K}  +
\vec{q}_{\theta}$  ($\forall  \theta$ such  that  $a_{\theta} \ne  0$)
represent   unfolded  states   corresponding   to  the   non-primitive
wavevector  $\vec{K}$.  Note  that  these may  lie  outside the  first
primitive Brillouin zone, in which  case, they need to be shifted back
in using an appropriate primitive reciprocal lattice vector.

We  now  clarify  an  issue  related  to  determining  the  values  of
$\tilde{C}'^{\mu   \varsigma  n}$   from  the   eigenvectors   of  the
non-primitive    version    of    palindromic    eigenvalue    problem
(\ref{eq_SchrodingerMatrixPrimitive}).       Note      first      that
$\mathtt{constant} \times [C_t]$ is as good an eigenvector as $[C_t]$,
where $\mathtt{constant}$ is in any complex number independent of $\mu
\varsigma n$.  Thus, the eigensolver can be thought of as returning an
eigenvector,  $\mathtt{constant} \times  [C_t]$, normalized  such that
$\sum_{\mu \varsigma  n} |\mathtt{constant} \times[C_t]_{\mu \varsigma
n}|^2 = 1$.  Now, from (\ref{eq_C_CPrimed}), (\ref{eq_CPrimeTilde}) we
have
\begin{eqnarray}
\label{eq_Cmystery}
\tilde{C}'^{\mu \varsigma n}(\vec{K}^{\perp}) = 
e^{\iota \vec{K}^{\parallel} \cdot \vec{\gamma}_n} 
\times \underbrace{
\sqrt{S_{NP}(\vec{K}^{\perp})}
e^{- \iota \vec{K}^{\perp} \cdot t \vec{f}_1}
\times 
[C_t]_{\mu \varsigma n}
}_{\mathrm{eigenvector}}
\end{eqnarray}
Choosing  $\mathtt{constant}   =  \sqrt{S_{NP}(\vec{K}^{\perp})}  e^{-
\iota   \vec{K}^{\perp}  \cdot   t  \vec{f}_1}$,   we   can  associate
$\tilde{C}'^{\mu  \varsigma  n}$ with  the  eigenvector  returned by  the
solver,   after    scaling   individual   rows    are   by   $e^{\iota
\vec{K}^{\parallel}    \cdot    \vec{\gamma}_n}$,    as    shown    in
(\ref{eq_Cmystery}). Since $\vec{K}^{\parallel}$ is real, this ensures
that $\sum_{\mu \varsigma n} |\tilde{C}'^{\; \mu \varsigma n}|^2 = 1$.

Finally, a  we would like to  comment on the  possible implications of
this  work  to  determine  the  complex  bandstructure  of  disordered
materials. The  supercell method computes  energy bands using  a large
non-primitive supercell, and unfolds these onto a fictitious primitive
small-cell.   This   supercell  is  non-primitive   w.r.t  $\vec{u}_2,
\vec{u}_3$  (i.e.   $\vec{f}_2  =   N_2  \vec{u}_2,  \vec{f}_3  =  N_3
\vec{u}_3$ for integers $N_2, N_3 > 1$).  As mentioned earlier, a careful
choice of linearization scheme  and eigensolver is essential to obtain
useful results. Systems  with disorder  can  be thought  to have  a
spread  in their  $E(\vec{k})$ dispersion  -- i.e.  at  any $\vec{k}$,
there  are states with  energies within  an interval  given by  a mean
energy  $\bar{E}$,  and  a  deviation  $\delta  E$  about  this  mean.
Equivalently, at any  energy $E$, each complex band  can be thought of
having  a   mean  $\bar{\vec{k}}^{\perp}(E)$  and   a  spread  $\delta
\vec{k}^{\perp}(E)$.   The  central idea  of  the supercell  technique
applied to real bands is to extend the summation in (\ref{eq_Psi_psi})
so that a supercell state $\ket{\Psi_p(\vec{K})}$ with energy $E_p$ is
expressed in terms  of $N_{O\, cell}$ small cell  states with energies
$E_{\eta}$ as
\begin{eqnarray}
\label{eq_Psi_psi_Disorder_Boykin}
\ket{\Psi_p(\vec{K})} =
\sum_{\eta = 1}^{N_{O \, cell}}
\sum_{\theta = 1}^{\mathcal{N}_c}
a_{\eta, \theta ; p}
\ket{\psi_{\eta}(\vec{K} + \vec{q}_{\theta})},
\end{eqnarray}
where  $N_{O  \,  cell}$ refers  to  the  number  of orbitals  in  the
small-cell, and hence is the  number of small-cell energy bands at any
given  $\vec{k}$.  It  is  reasonable  to expect  that  the  supercell
technique,  when  extended to  compute  the  complex bandstructure  of
disordered materials will similarly involve a summation of states with
different energies.   The invariance  of $\mathcal{M}$ on  energy will
hence  be useful  in simplifying  computation. The  details of  such a
computation are beyond the scope of the present work, and could be the
subject of further study.


\begin{figure}[t]
 \centering	
 \includegraphics[scale=1]{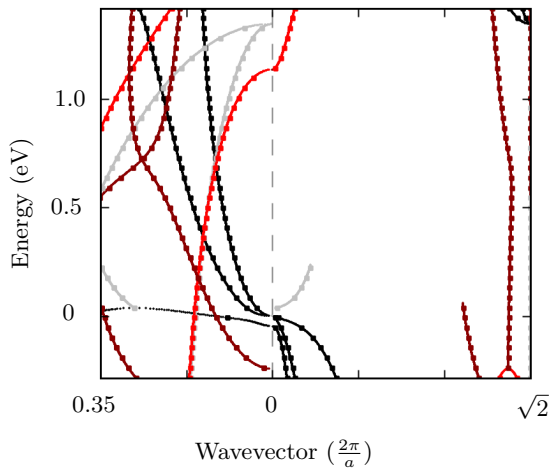}
 \caption{Complex  bandstructure  of  Silicon (lattice  constant  $a$)
along   $[110]$.    Tight    binding   parameters   are   taken   from
\cite{Boykin_PRB_2004}.   Real and imaginary  parts of  the wavevector
are shown  on the  right and left  panels respectively. Pure  real and
pure imaginary  bands are in black,  red whereas complex  bands are in
grey,    dark    red     for    $\vec{k}^{\parallel}    =    \vec{0}$,
$\vec{k}^{\parallel}   =    (0,0,   0.84   \times   2    \pi   /a   )$
respectively. Lines represent  results using primitive vectors whereas
the  filled  squares  represent  results using  non-primitive  vectors
followed by zone unfolding.}
 \label{fig_Si110}
\end{figure}

\section{Application and Summary}
\label{sec_Discussion}
Transport along  the $[110]$  direction leads to  a quartic  GPEP when
working  with primitive  vectors.   On the  other  hand, the  smallest
non-primitive unit  cell such that $\vec{f}_1 \parallel  \vec{n}$ is a
double  cell ($\mathcal{N}_c  =  2$).  Thus,  there  are two  possible
primitive  wavevectors that each  non-primitive wavevector  can unfold
onto.   Figure \ref{fig_Si110} compares  the complex  bandstructure of
Silicon along  the $[110]$ direction, obtained using  a primitive cell
with  that obtained  using this  non-primitive cell,  followed  by our
zone-unfolding  procedure.  Tight  binding parameters  are  taken from
   \cite{Boykin_PRB_2004}.     Two   different   value   of
$\vec{k}^{\parallel}$ are considered, corresponding to $\vec{k}$ paths
through $(0,0,0)$ (valence band maximum)  and $(0,0, 0.84 \times 2 \pi
/a )$ (one of the $\Delta$ conduction valleys).  The two methods yield
identical results.  Further,  Table \ref{tab_measure} demonstrates the
invariance of the  measure $\mathcal{M} = \sum_{\theta}|a_{\theta}|^2$
on energy $E$ ensured by the inclusion of the factor $\Lambda$.

In conclusion, we have derived  a unified method of unfolding real and
complex bands in a nearest neighbour tight-binding scheme.  This method
reduces   to   the    unfolding   method   available   in   literature
\cite{Boykin_PRB_2005},  for  the  case  of real  bands.   Using  this
unfolding method, complex bands  along any general transport direction
$\vec{n}$ can be  computed by the solution of  a generalized quadratic
eigenvalue problem,  using a non-primitive unit  cell.  This overcomes
the  difficulties  regarding the  solution  of generalized  polynomial
eigenvalue  problems of large  order, that  may result  when computing
complex bands  using primitive  cells for general  $\vec{n}$. Finally,
our method ensures an energy  invariant measure for the projections of
the non-primary wavefunction onto all candidate primary wavefunctions.
This  invariance will  be  important for  computing  complex bands  of
disordered      materials     using     a      supercell    approach
\cite{Boykin_JPhysCondMatter_2007}.

\begin{table}[t]
  \caption{Effect of factor  $\Lambda$ (\ref{eq_final}) on the measure
$\mathcal{M}  =  \sum_{\theta}|a_{\theta}|^2$  for states  having  the
smallest  values of  $|Im(k^{\perp}(E))|$  and $\vec{k}^{\parallel}  =
(0,0, 0.84  \times 2 \pi /a  )$.  $\mathcal{M}_{old}$ is  the value of
$\mathcal{M}$  setting  $\Lambda =  1$,  corresponding  to the  result
provided    by    the     real    band    unfolding    algorithm    of
\cite{Boykin_PRB_2005}. $\mathcal{M}_{new}$  corresponds  to the
measure as  computed by the modified unfolding  algorithm described in
this work.}
  \begin{indented}
    \item[] \begin{tabular}{ccccc}
      \br
      $E \; (eV)$ &  
      $K^{\perp} \; ( 2 \pi / a )$  & 
      $k^{\perp} \; ( 2 \pi / a )$  &
      $\mathcal{M}_{old}$ &   
      $\mathcal{M}_{new}$ \\
      \mr
      $0.05$        & 
      $-0.116 + 0.126\iota$ & 
      $1.299 + 0.126\iota$ &
      $1.272$    &
      $1.0$ \\

      $0.8 $ &
      $0.103 \iota$  &
      $0.103 \iota$ &
      $1.224 $&
      $1.0$ \\    
      \br  
    \end{tabular}
  \end{indented}
  \label{tab_measure}
\end{table}

\ack
The authors  wish to thank Dr.  S. E.  Laux, IBM  USA, for stimulating
discussions and access to his complex bandstructure code and Dr. Rajan
Pandey,  Dr.  Samarth  Aggarwal (IBM  India), G.  Vijaya Kumar  (IIT
Madras) for helpful suggestions. A. Ajoy wishes to thank IBM India for
financial support.

\appendix
\section*{Appendix}
\section{Primitive wavefunction in terms of atomic
    orbitals}

Replacing $\vec{f},  t, \vec{R}, n,  N, \Xi, \Psi, C'$  describing the
non-primitive  wavefunctions with  $\vec{u},s,  \vec{\rho}, m,M,  \xi,
\psi,          c'$           respectively          in          Section
\ref{sec_WavefnIntermsofAtomicOrbitals},    we   have    the   primary
wavefunction
\begin{eqnarray}
\label{eq_psi:1}
\nonumber \fl \ket{\psi(\vec{k}^{\perp}, \vec{k}^{\parallel})} 
 =  \frac{1}{\sqrt{M_{\parallel}S_{P}(\vec{k}^{\perp})}} 
\sum_{\mu \varsigma m}
\sum_{j}^{(M_{\parallel})}
&\sum_{s = 0}^{M_1 - 1} 
\tilde{c}'^{\; \mu \varsigma m}(\vec{k}^{\perp}) \times \\
& e^{\iota  \vec{k}^{\perp} \cdot s\vec{u}_1 }
e^{\iota \vec{k}^{\parallel} 
\cdot (\vec{\rho}^{\parallel}_j + s\vec{u}_1)}
\ket{\mu,\varsigma; \vec{\rho}^{\parallel}_j + s\vec{u}_1 +
  \vec{\nu}_m}
\end{eqnarray}
In going  from (\ref{eq_psi:1}) to  (\ref{eq_psi:2}), we use  the fact
that by construction , $\vec{k}^{\perp} \cdot \vec{\rho}^{\parallel}_j
= 0$.  Hence, the exponent in (\ref{eq_psi:1}) is simplified as
\begin{eqnarray}
\label{eq_kparallel.kperp:k}
\vec{k}^{\perp} \cdot s\vec{u}_1  + 
\vec{k}^{\parallel} \cdot (\vec{\rho}^{\parallel}_j + s\vec{u}_1)
= \vec{k} \cdot \vec{\rho}_{j'}, 
\end{eqnarray}
where $\vec{\rho}_{j'} = (\vec{\rho}^{\parallel}_j + s\vec{u}_1)$. The
double  summation   in  (\ref{eq_psi:1}),  $\sum_{j}^{(M_{\parallel})}
\sum_{s  = 0}^{M_1 -  1} \equiv  \sum_{j'}^{(\mathcal{N}_{P})}$ (where
$\mathcal{N}_{P}  =  M_{\parallel}M_{1}$   refers  to  the  number  of
primitive lattice points) and the $'$ can finally be dropped from $j'$.

Imposing     the     conditions    that     $\ip{\psi(\vec{k}^{\perp},
\vec{k}^{\parallel})}  {\psi(\vec{k}^{\perp},  \vec{k}^{\parallel})} =
1$   and    $\sum_{\mu   \varsigma   m}    |\tilde{c}'^{\;   \mu   \varsigma
m}(\vec{k}^{\perp})|^2 = 1$, we get
\begin{eqnarray}
\label{eq_S_P}
\nonumber S_{P}(\vec{k}^{\perp}) &=& \sum_{s = 0}^{M_1 - 1}
e^{-s \beta }, 
\mbox{ where } \beta = 2 Im(\vec{k}^{\perp} \cdot \vec{u}_1) \\
& = & 
\cases{  M_1,  &  if  $\beta = 0$ \\
         \frac{1 - e^{-\beta M_1}}{1 - e^{-\beta}}, 
         & if $ \beta  \ne 0.$
      }   
\end{eqnarray}

\section*{References}
\providecommand{\newblock}{}

\end{document}